\documentstyle[prl,aps,epsf,floats]{revtex}

\begin{document}


\twocolumn[\hsize\textwidth\columnwidth\hsize\csname@twocolumnfalse%
\endcsname

\title{Non-perturbative Effects Near the Upper Critical Field in
Type--II Superconductors}

\author{Safi R. Bahcall}
\address{Department of Physics, University of California, 
Berkeley CA 94720}
\maketitle
          
\begin{abstract}

The properties of a conventional type-II superconductor near the upper
critical field are usually calculated using a perturbative expansion
in the strength of the order parameter.  Here we show that perturbation
theory breaks down near the upper critical field and that this
breakdown leads to unusual behavior near the phase transition.

\vspace*{.1in}

\end{abstract}


]

\def\bottomfraction{.9}
\def\textfraction{.1}

Standard microscopic calculations of the properties of type-II
superconductors near the upper critical field rely on two assumptions:
1) a perturbative expansion in the strength of the superconducting
order parameter and 2) weak magnetic field.  These calculations yield
an upper critical field $H_{c2}(T)$ close to what is observed
experimentally \cite{gorkov59,fetter}.  More complete solutions to the
Gor'kov equations, which do not rely on the weak magnetic field
assumption, yield low temperature divergences in $H_{c2}(T)$
\cite{gruenberg,norman,quanreg} which have not been observed.  In
this paper we show that the origin of these divergences is the
breakdown of perturbation theory, and that the full, non-perturbative
solution to the Gor'kov equations describes tails of residual 
superconductivity above $H_{c2}$.

The microscopic description of type II superconductors is based
on Gor'kov's extension of the BCS theory
\begin{equation}
\label{gorkovh}
 {\cal H}=\int\! d{\bf r} \,c^\dagger_{{\bf r}\sigma} {\cal H}_0({\bf r}) 
 c^{\vphantom{\dagger}}_{{\bf r}\sigma} - 
{1\over 2}\,V_0 \,\Omega \int\! d{\bf r}\, c^\dagger_{{\bf r}\sigma}
 c^\dagger_{{\bf r}\sigma'} c^{\vphantom{\dagger}}_{{\bf r}\sigma'} 
 c^{\vphantom{\dagger}}_{{\bf r}\sigma} ,
\end{equation}
which describes a short-range attractive ineraction of strength
$V_0>0$ for particles near the Fermi energy, $|{E}-{E_F^{}}|<{E_c}$.
${\cal H}_0$ is the bare Hamiltonian ${\cal H}_0$ $\equiv$ $(1/2m)$
$\left( i{\bf\nabla} - e{\bf A}/c\right)^2-{E_F^{}} $,
$c^\dagger_{{\bf r}\sigma}$ are the electron creation
operators, and $\Omega$ is the system volume.  The variational approach
to this problem is to solve for the eigenstates of the two-body
Hamiltonian
\begin{equation}
\label{twobodyh}
 {\cal H}' = \int \!d{\bf r} 
\; \, \Psi_{\bf r}^\dagger \;
 \left[ \begin{array}{cc} {\cal H}_0({\bf r}) & \Phi({\bf r}) 
 \\ \Phi^\ast({\bf r})  & -{\cal H}_0^\ast({\bf r}) \end{array} \right]
\; \Psi_{\bf r}^{\vphantom{\dagger}} \ 
\end{equation}
where $\Psi_{\bf r}^\dagger \equiv 
\big[ \, c^\dagger_{{\bf r}\uparrow} \; 
c^{\vphantom{\dagger}}_{{\bf r}\downarrow} \, \big]$.
Minimizing $\langle {\cal H} \rangle$ 
leads to the self-consistency condition
\begin{equation}
\label{selfcons}
 \Phi({\bf r}) = -V_0\,\Omega\,
 \langle c^{\vphantom{\dagger}}_{{\bf r}\uparrow} 
c^{\vphantom{\dagger}}_{{\bf r}\downarrow}  \rangle \ .
\end{equation}
The standard result for the upper critical field $H_{c2}$ follows from
evaulating Eq. (\ref{selfcons}) perturbatively in $\Phi({\bf r})$.
Both sides of this equation are linear to lowest order so setting the
resulting coefficients equal gives the condition which determines
$H_{c2}$.  Although this is usually done using Green's functions, 
here we consider the wavefunctions directly.

First we separate the magnitude of the order parameter 
\begin{equation}
\Phi({\bf r}) \ \equiv\ \phi\; f({\bf r})
\end{equation}
where $\phi$ is real and positive and $f({\bf r})$ is normalized so that
${1\over \Omega}\int |f({\bf r})|^2\,d{\bf r} =1$.  Next, we 
write the eigenstates of the bare Hamiltonian as 
\begin{equation}
{\cal H}_0\; \psi_\alpha({\bf r}) \ = 
\ \varepsilon_\alpha\; \psi_\alpha({\bf r})\ 
\end{equation}
and define new quasiparticle operators by rotating to this basis:
$d_{\alpha\sigma}^\dagger = \int\! d{\bf r}\; \psi_\alpha({\bf r}) 
c_{{\bf r}\sigma}^\dagger$.
We can then define a pairing matrix
\begin{equation}
A_{\alpha\beta} \ \equiv \ \int\!\!d{\bf r} \; f({\bf r})\,\psi_\alpha^{\vphantom{\dagger}}
({\bf r}) \,\psi_\beta^{\vphantom{\dagger}}({\bf r}) \ ,
\end{equation}
so that the variational two-body Hamiltonian  is
\begin{equation}
\label{gentbh}
{\cal H}' =  \Psi^\dagger\;      \left[ \begin{array}{cc} \varepsilon
  & \phi\, A \\
  \phi^\ast A^\dagger & -\varepsilon \end{array}  \right]\,  \Psi\ ,
\end{equation}
where $\Psi$ is the vector of quasiparticle operators 
$\Psi^\dagger \equiv  \big[\, \cdots\, d_{\alpha\uparrow}^\dagger \cdots
\; \cdots\, d_{\beta\downarrow}^{\vphantom{\dagger}}\cdots \, \big]$ 
and $\varepsilon$ is the diagonal matrix of bare eigenvalues
$|\varepsilon_\alpha|<E_c$.

Eq. (\ref{gentbh}) is the general form of the pairing Hamiltonian that
occurs when there is no translation or time-reversal symmetry
restricting the states which can be paired.  If $\phi$ is small and we
can use perturbation theory, the variational energy difference between
the superconducting and normal state, $\langle {\cal H} 
\rangle_S - \langle
{\cal H} \rangle_N$, that follows from Eqs. (\ref{selfcons}) and
(\ref{gentbh}) is, keeping the lowest two orders in $\phi$:
\begin{equation}
\label{delezero}
\Delta E_{\rm pert}^{}(H) = \phi^2 \, \left( \, {1\over V_0} - \sum_{\alpha\beta}
{|A_{\alpha\beta}|^2\over |\varepsilon_\alpha+\varepsilon_\beta|} 
\,\right) + b\, \phi^4 \ .
\end{equation}
This describes a continuous phase transition at the field $H$ at which the
quantity in parenthesis vanishes.  When the electron gas wavefunctions
are used, this yields the semiclassical result for $H_{c2}$ plus some
cutoff-dependent corrections \cite{bahcall}.

\begin{figure}[b]
\epsfysize=6cm\centerline{\epsfbox{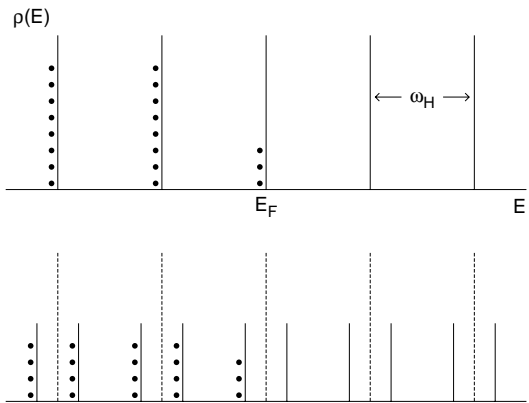}}
\vspace{6pt}
\caption{Top: Density of states of an ideal two-dimensional electron
gas.  At zero temperature the Landau level at the Fermi energy is
partially filled.  Bottom: The pairing 
splits each level into two levels separated by an energy $\propto \phi$, 
the magnitude of the order parameter.  Electrons in the partially
filled level contribute a net $O(\phi)$ term to the energy.}
\label{figllsplit}
\end{figure}

Non-perturbative effects are illustrated most simply for a
two-dimensional electron gas at zero temperature.  Bare energies are
Landau levels, multiples of the cyclotron energy $\hbar\omega_H^{}$.  The
Fermi energy is at the highest occupied Landau level, so the matrix
$\varepsilon$ in Eq. (\ref{gentbh}) is
\begin{equation}
\label{bareh}
\varepsilon = \hbar\omega_H^{} \left(\;  \begin{array}{*{7}{r@{\;\,}}}
   \makebox[2pt]{$\ddots$} &  &  &  &  &  &  \\[-1ex]
     & \makebox[1.8ex]{$-\!2$} &  &  & &  &  \\[-1ex]
     &  & \makebox[1.8ex]{$-\!1$} &  &  &  &  \\[-1ex]
     &  &  &  \makebox[1.3ex]{$0$} &  &  &  \\[-1ex]
      &  &  &  & \makebox[1.3ex]{$1$} &  &    \\[-1ex]
     &  &  &  &  & \makebox[1.3ex]{$2$} &  \\[-1ex]
     &  & &  &  &  &  \raisebox{-.6ex}[0pt]{\makebox[9pt]{$\ddots$}}
\end{array} \right) 
\end{equation}
Near the phase transition $\phi$ can be taken to be small compared to
$\hbar\omega_H^{}$, but before perturbation 
theory can be applied degenerate
levels must be diagonalized.  The pairing splits each Landau level
into two with energies $\varepsilon_m \pm \phi |A_{m,2N_f-m}|$.  For the
filled levels below the Fermi energy the number of occupied states
shifted up in energy equals the number shifted down in energy so there
is no net contribution to the energy 
from this splitting.  When the level right at the Fermi energy is
partially filled, however, more states will be shifted down in energy
(Fig.\ \ref{figllsplit}).  This will contribute a term linear in
$\phi$ to the energy difference between the superconducting and 
normal state, which then has the form:
\begin{equation}
\label{deleA}
\Delta E_{\rm 2D}^{}(H)\ =\ -\epsilon\,\phi + a(H)\,\phi^2 + O(\phi^3)\ .
\end{equation}
In the relevant weak field limit, $E_F^{} \gg \hbar\omega_H^{}$, and
assuming the lowest Landau level component for the order parameter, a
calculation for the electron gas yields $\epsilon=\alpha\, \rho_0\,
\hbar\omega_H^{}$ and $a(H)=\rho_0\, \ln(H/H_{c2}^0)$, where $\rho_0$
is the zero-field density of states at the Fermi energy and $H_{c2}^0$
is the perturbative result for the upper critical field.  Here
$\alpha\approx 0.67\, \nu/N_f^{1/4}$ where $\nu$ is the fraction of
the highest Landau level which is filled (or one minus this fraction
if it is more than half-filled) and $N_f$ is the total number of
filled levels $N_f=E_F^{}/\hbar\omega_H^{} =(k_F^{}\ell_H^{})^2/2\sim
10^3$--$10^4$ for typical materials.

\begin{figure}[b]
\epsfysize=6cm\centerline{\epsfbox{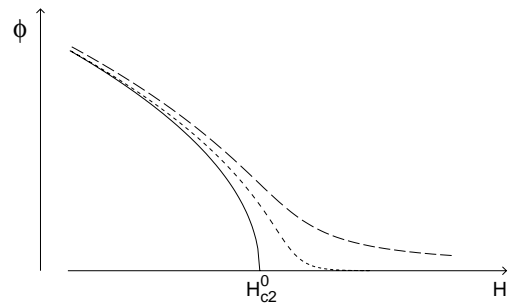}}
\vspace{6pt}
\caption{Solid line is the standard perturbative result 
for the order parameter
as a function of field near the transition: $\phi\propto (H_{c2}-H)^{1/2}$.
Two-dimensional residual superconductivity at $T=0$, Eq.
(\protect{\ref{deleA}}), yields rapid oscillations about the top
dashed line; the three-dimensional form, Eq. (\protect{\ref{deleB}}), 
yields the middle dashed line.}
\label{figphiofh}
\end{figure}

Minimizing Eq. (\ref{deleA}) for the energy difference yields the
result shown in Fig.\ \ref{figphiofh}: the order parameter develops a
``tail'' of persisting superconductivity.  In the limit of small
$\epsilon$, the magnitude of the order parameter in the tail is 
$\phi = \epsilon/2 a(H)$.  Since $\epsilon$ depends
on the filling factor of the highest level, $\phi$ oscillates with
magnetic field at a frequency $\delta H/H_{c2} \approx 1/N_f$ about
the upper dashed line in Fig. \ref{figphiofh}.

This form of residual superconductivity depends on the exact
degeneracy of the Landau level poles of a non-interacting two-dimensional
electron gas in a magnetic field.  In real materials
interactions will broaden these levels.  When the broadening is larger
than $\sim\epsilon/a(H)$, there will no longer be a term linear in $\phi$
in the energy difference and this effect will be suppressed.  

It is believed that the effect of impurities in two dimensions at zero
temperature is to cause extended states to appear only at discrete
energies, as evidenced by the quantum Hall effect
\cite{laughlin,halperin}.  To the extent to which the most relevant
interactions are those between the extended states, pairing
will split these levels as in Fig. \ref{figllsplit} and cause the
two-dimensional residual superconductivity to occur in
superconductors at temperatures low enough and materials clean enough
that the quantum Hall effect could be resolved in the normal state.
Such conditions have not yet been achieved in thin metallic films, 
but may eventually be.

Although the two-dimensional form of residual superconductivity
depends on large degeneracies in the spectrum, a weaker form
occurs more generally, when the normal-state spectrum is flat.
To see where this comes from, consider first the zero-field BCS
theory.  The difference between the ground state energy of the
superconductor and of the normal state can be written, for a uniform
gap $\Delta$,
\begin{equation}
\label{deleBCS}
\Delta E_{BCS}^{}\, = \,  {\Delta^2\over V_0} -
2\,\rho_0 \int_0^{E_c} \! dE\; \Big(
\sqrt{E^2 + \Delta^2} - E \Big)\, .
\end{equation}
The first term is from the expectation value of the interaction term
in $H$ and the second is the shift in the electronic energy levels.
The normal state spectrum is assumed flat in the range $E_c$
around the Fermi energy.

Even for small $\Delta$ compared
to $E_c$, perturbation theory can not be used to evaluate the integral
in the second term of Eq. (\ref{deleBCS}).
The  exact answer is
${1\over 2} \Delta^2\big[\ln(2E_c/\Delta) + 
{1\over 2} + O(\Delta^2/E_c^2)\,\big]$: 
the $\Delta^2 \ln\Delta$ term is non-perturbative.  In
the BCS case this term exactly cancels the $\Delta^2/V_0$ 
term, leaving the net
condensation energy ${1\over 2} \rho_0\Delta^2$.  In a magnetic 
field, the balance
between the $O(\phi^2)$ terms determines the perturbative upper critical
field, as in Eq. (\ref{delezero}), and a net $\phi^2\,\ln\phi$ term
remains:
\begin{equation}
\Delta E_{\rm 3D}^{}(H) = \epsilon\,\phi^2\,\ln\phi
+ a(H)\,\phi^2 + O(\phi^3)\ .
\label{deleB}
\end{equation}
This form for the energy difference can be derived for the
three-dimensional electron gas in the formal limit, as before,
$\phi\ll\hbar\omega_H^{}$.  The additional complication is that the bare
Hamiltonian does not have the simple structure of Eq. (\ref{bareh});
the energies are offset by an amount $\delta_z$ which depends on the
$z$-momentum: $\varepsilon_n(k_z)= \hbar\omega_H^{}
 (n+{1\over 2}) + k_z^2/2m -
E_F^{} \equiv \hbar\omega_H^{}(n-N_f + \delta_z)$.  Integrating with respect to
$\delta_z$ those levels within $\hbar\omega_H^{}$ of the Fermi energy gives an
integral of the form which occurs in Eq. (\ref{deleBCS}).  The result
is Eq. (\ref{deleB}) with $\epsilon \approx 0.22 \,\rho_0 /N_f^{1/2}$ and
$a(H)=\rho_0\,\ln(H/H_{c2}^0)$.  

\begin{figure}[b]
\epsfysize=5cm\centerline{\epsfbox{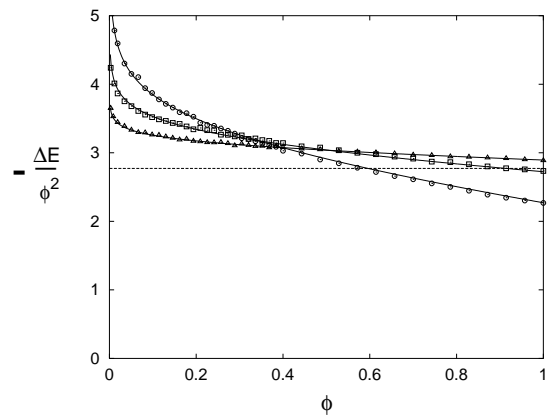}}
\vspace{20pt}
\caption{Change in ground state energy ${\it vs.}$ order parameter 
strength $\phi$ for the random matrix ensemble
defined in the text.  Horizontal dashed line is the perturbative
$N\rightarrow \infty$ result.
Points are for $N=5$, $10$, $20$, (\protect{$\circ$, 
$\protect\raisebox{1mm}{\protect\framebox[2mm]{}}\,$, $\Delta$}) 
and solid lines
are fits to the form of Eq. (\protect{\ref{deleB}}): 
\protect{$\Delta E/\phi^2 = \alpha
\,\log\phi - \beta + \gamma\,\phi$}.
Both $\alpha$ and $\gamma$ decrease as $\sim 1/N$.  $10^4$ matrices were
used for each point.     }
\label{figrandom}
\end{figure}

The form of Eq. (\ref{deleB}) is not sensitive to the assumption of a
non-interacting electron gas in a magnetic field.  If we suppose that
1) the spectrum in the normal state is flat, and 2) the radiply varying
matrix elements that enter Eq. (\ref{gentbh}) are random, we obtain
the results shown in Fig.\ \ref{figrandom}.
The diagonal matrix $\varepsilon$
is taken from a uniform distrubition 
$-N/2<\varepsilon_i<N/2$, corresponding
to the flat normal state spectrum.  The $N\times N$ matrix
$A_{\alpha\beta}$ is taken to be a complex symmetric matrix normalized
to ${\rm Tr} A^\dagger A = N$ with individual elements Gaussian
distributed.  (The $\phi\rightarrow 0$ results are not very sensitive
to this choice of distribution, or to a more realistic approximation
where $A_{\alpha\beta}$ falls off exponentially with
$|\varepsilon_\alpha-\varepsilon_\beta|$.)  
We calculate the shift in the ground state
energy $\Delta E$, divide by $\phi^2$, and examine the limit as
$\phi\rightarrow 0$.  If the result were purely perturbative,
corresponding to Eq. (\ref{delezero}), we would see a constant
limit.  Instead, a logarithmic upturn is clearly visible, corresponding
to Eq. (\ref{deleB}).

Whether the non-perturbative effects will be visible in a given
material depends on the number of states $N$ that are mixed by the
pairing.  In the electron gas case, the symmetry of the vortex lattice
generates a conserved momentum and $N\sim \sqrt{N_f}$.  The effects of
disorder in a superconductor, when time-reversal symmetry has been
explicitly broken by the presence of a magnetic field, are not well
understood. (The semiclassical Abrikosov-Gor'kov description is not
sufficient to address the effects discussed here.)  If the number of
states being paired becomes macroscopic then non-perturbative effects
will be suppressed.  If the effective number of states being paired
remains small, the number of states inside a volume determined by the
scattering length rather than the coherence length, then
non-perturbative effects will remain and be enhanced.

A further interesting possibiltity is that the random matrix model
described above is a useful model for the spectrum of a granular
superconductor \cite{jaeger} in a field, where pairing takes place
within a grain and is in the mesoscopic regime.  In that case $N$ is
the number of states in a grain being paired, and can be as low as
$10^2$.

At fields far above the conventional $H_{c2}$, minimizing
Eq. (\ref{deleB}) for the order parameter strength gives
$\phi\sim\exp(-a(H)/\epsilon)$.  The order parameter becomes
exponentially suppressed, as shown in Fig. \ref{figphiofh}.  When
$\phi$ is sufficiently small the superconductivity will
become disfavored due to normal-state Pauli paramagnetism or
pair-breaking effects, such as magnetic impurities, which are not
included in the idealized Hamiltonian in Eq. (\ref{gorkovh}).  It is
an interesting possiblity, however, that more exotic order parameter
configurations \cite{fulde} would allow the residual state to persist
even in the presence of such effects.

At finite temperature, the quantity to be minimized is the Gibbs free
energy.  As in the BCS case, the energy difference $\Delta E(T) =
\phi^2/V_0 - 2\int\,dE\,E\,\delta\rho(E) \tanh(E/2T) $, where
$\delta\rho$ is the change in the density of states, and the entropy
is given by the Fermi gas expression.  The factor of $\tanh(E/2T)$
cuts off the $\phi^2/E^2$ rise of $\delta\rho(E)$ at low energies
which generated the $\phi^2 \ln\phi$ term.  Initial numerical results
indicate that non-perturbative effects remain in the $O(\phi^3)$
contribution to the free energy.  This would have interesting implications
for behavior near the phase transition.

Usually an $O(\phi^3)$ term in the free energy
\begin{equation}
\label{myg}
G\ =\  a\,t\,\phi^2\ +\  \gamma\,\phi^3\ +\ b\,\phi^4 \ +\ \ldots\ ,
\end{equation}
where $t\equiv T-T_c(H)$, signals a first order phase transition.  
Defining
\begin{equation}
\phi^\ast\ \equiv \gamma/b\quad , \quad t^\ast\ \equiv\ 
9\gamma^2/32 a b\ ,
\end{equation}
and keeping terms up to $O(\phi^4)$, we have that a second minimum
appears in $G(\phi)$ for temperatures below $t^\ast$ and becomes
energetically favorable compared to the $\phi=0$ solution when
$t<t_1=(8/9)t^\ast$.  More precisely, the two non-zero extrema of $G$
which occur at temperatures below $t^\ast$ are at
\begin{equation}
\phi_{\pm}=  {3\over 8}\;\phi^\ast
\left(-1\pm \sqrt{1-(t/t^\ast)}\right)\ .
\end{equation}
For negative $\gamma$, $\phi_{+}$ is the local minimum which becomes
energetically favorable below $t_1$, causing a first-order transition.
For positive $\gamma$, the local minimum $\phi_{-}<0$.  $\phi$,
however, is the magnitude of the order parameter and must be positive,
so this solution is not allowed.  What happens instead, if $\gamma$ is
positive, is that there is a continuous transition at $t=0$ as the
$\phi_+$ solution becomes an allowed minimum.  Near this transition,
for $|t|\ll t^\ast$, expanding the square root yields
\begin{equation}
\phi \approx {3\phi^\ast \over 16 t^\ast} \, |t|\ .  
\end{equation}
The scale for this deviation from the familiar mean-field behavior
$\phi \propto \sqrt{|t|}$ is set by $t^\ast$: for $|t|\gg t^\ast$ the
square-root behavior is recovered.  The same behavior occurs if the
field is varied at fixed temperature; $t$ is replaced by $h\equiv
H-H_{c2}(T)$ and the result is $\phi\propto |h|$ for $|h|\ll h^\ast$.
Although the models considered here give a negative $\gamma$, favoring
the continuous transition, the possibility that additional physics
enters at sufficiently small $\phi$, and generates a positive
$\gamma$, can not be ruled out.

The deviations from standard mean field behavior described here would
be difficult to see in resistivity measurements, the most common
measurements done near $H_{c2}$ on conventional materials.  They would
be more likely to be seen in sensitive heat capacity, magnetization,
or tunneling measurements.  The form Eq. (\ref{myg}) for the free
energy gives $G\propto (T-T_c)^3$ near the continuous transition (for
$t\ll t^\ast$), instead of the Ginzburg-Landau behavior $G\propto
(T-T_c)^2$, and leads to a kink in the heat capacity rather than an
abrupt discontinuity.  The magnetization $M\propto \phi^2$ would vary
as $(T_c-T)^2$ rather than $(T_c-T)$.  In the perturbative mean field
theory, the spectrum near $H_{c2}$ changes by $O(\phi^2)$.
Non-perturbative effects arise from states close to the Fermi energy
being strongly affected by the pairing even as $\phi\to 0$.  This
leads to a zero-bias anomaly in the spectrum: a peak near the Fermi
energy, a $\phi^2/E^2$ rise, at the center of which is is a sharp
spike giving $\rho(E_F^{})=0$.

Non-perturbative effects appear at scales much larger than the
deviations from mean field behavior expected from phase fluctuations.
The phase fluctuations which cause vortex lattice melting are believed
to become important on the scale of critical phenomena, as estimated
by the Ginzburg criterion \cite{lobb}.  This scale is
$(T_G^{}-T_c)/T_c \alt 10^{-16}$ in zero magnetic field 
and $\alt 10^{-7}$ at $H=10\, T$ for typical parameters of bulk
conventional superconductors, although the scale may be much larger in
high-$T_c$ materials \cite{lobb}.  The scale for the deviations
described here is $1/\sqrt{N_f}$ which can be $1\%$--$10\%$ for
conventional supercondcutors.

I would like to thank R.~B.~Laughlin for support from NSF Grant
No. DMR--9120361.  This work was also supported by a Miller Research
Fellowship from the Miller Institute for Basic Research in Science.

\smallskip

\end{document}